\documentclass[conference]{IEEEtran}

\usepackage[cmex10]{amsmath}
\usepackage[pdftex]{graphicx}
\usepackage{latexsym}
\usepackage{amssymb}
\usepackage{bm}
\usepackage{psfrag}
\usepackage{algorithm}
\usepackage{algorithmic}

\usepackage[pdftex]{graphicx,xcolor}
\usepackage{cite}

\newtheorem{def.}{Definition}
\newtheorem{th.}{Theorem}
\newtheorem{le.}{Lemma}
\newtheorem{ex.}{Example}
\newtheorem{re.}{Remark}

\renewcommand{\vec}[1]{\boldsymbol{#1}}

\begin{document}

\sloppy

\title{
Construction and Encoding Algorithm for Maximum Run-Length Limited Single Insertion/Deletion Correcting Code
}

\author{
\IEEEauthorblockN{Reona Takemoto and Takayuki Nozaki}
\IEEEauthorblockA{
Dept. of Informatics, Yamaguchi University, JAPAN\\
Email: \{a029vbu,tnozaki\}@yamaguchi-u.ac.jp
}
}
\maketitle

\begin{abstract}
  Maximum run-length limited codes are constraint codes used in communication and data storage systems.
  Insertion/deletion correcting codes correct insertion or deletion errors
  caused in transmitted sequences and are used for combating synchronization errors.
  This paper investigates the maximum run-length limited single insertion/deletion correcting (RLL-SIDC) codes.
  More precisely, we construct efficiently encodable and decodable RLL-SIDC codes.
  Moreover, we present its encoding algorithm and show the redundancy of the code.
\end{abstract}

\section{Introduction}
The error control techniques play an important role to realize the reliable communication systems and data storage systems.
Many communication systems and data storage systems employ two types error control techniques, called error correcting codes and constraint codes.
The error correcting codes recover the errors caused in the transmitted sequences.
The constraint codes give the sequence which is suitable for the specific communication/storage requirements \cite{immink2004codes}.

Synchronization errors cause symbol insertions and symbol deletions in the transmitted sequences.
To combat such errors, many insertion/deletion correcting codes have been constructed.
In the construction of insertion/deletion correcting codes, there are mainly three approaches, namely, number-theoretic approach, probabilistic approach, and combinatorial approach.
In the number-theoretic approach, the codes are defined by single or multiple congruences \cite{VTcode,qVTcode,Levenshtein1966code,bibak2018weight,nozaki2019bounded,nozaki2020weight}.
In general, the number-theoretic codes are efficiently decodable and correct a fixed number of insertions/deletions.
In the probabilistic approach, the codes are decoded by message passing algorithms and can recover the insertions/deletions caused from statistical channel models \cite{davey2001reliable,koremura2020insertion,shibata2020design,shibata2020concatenated}.
By the combinatorial approach, we can obtain the code with large number of codewords.
However, in general, the most of codes constructed by combinatorial approach are not efficiently decodable \cite{mercier2010survey}.
In this paper, we focus on the number-theoretic codes.

Run length of a sequence is the number of the repetition of the same symbols.
The $r$-maximum run-length limited ($r$-RLL) code is a constraint code satisfying the maximum run-length of a sequence is smaller than or equal to $r$.
It is widely used for the communication and data storage systems, especially, DNA storage system \cite{yazdi2015dna}.

DNA storage system attracts attention as a future storage system, due to the longevity and high information density.
It is reported that the DNA storage system should satisfy the maximum run-length limited and GC-balanced constraints \cite{ross2013characterizing,heckel2019characterization}.
Immink and Cai constructed such constraint code \cite{Immink2020properties}.
Chee et al.\ gave efficiently encodable GC-balanced code correcting single insertion/deletion/substitution (IDS) \cite{chee2019linear}.
Cai et al.\ presented efficiently encodable $r$-RLL code correcting single insertion/deletion/substitution (IDS) \cite{cai2020efficient}.
However, the code given in \cite{cai2020efficient} suffers long maximum run-length $r$.

This paper constructs efficiently encodable/decodable $r$-RLL codes correcting single insertion/deletion with small $r$.
We call such code maximum run-length limited single insertion/deletion correcting (RLL-SIDC) code.

RLL-SIDC codes are also used for the construction of burst-insertion/deletion correcting codes.
Schoeny et al.\ \cite{schoeny2017codes} constructed binary $b$-burst insertion or deletion correcting codes, which correct any consecutive insertion or deletion error of length exactly $b$.
Non-binary $b$ burst insertion or deletion correcting codes are constructed in \cite{schoeny2017nonbinary, saeki2018}.
Construction of these codes uses interleaving of codewords,
i.e., matrix representation of codewords.
The first row of the matrix representation employs a RLL-SIDC code, e.g., RLL-VT code \cite{schoeny2017codes}.
The other rows employ bounded single insertion/deletion correcting (BSIDC) codes, e.g., shifted VT codes \cite{schoeny2017codes}, odd coefficient codes \cite{nozaki2019bounded}, or exponential coefficient codes \cite{nozaki2019bounded}.
Nowadays,
Lenz and Polyanskii proposed efficient binary codes that correct a $b$ or less burst insertion or deletion error \cite{lenz2020optimal}.

Any encoding algorithm has not been proposed to these burst insertion/deletion correcting codes \cite{schoeny2017nonbinary, saeki2018, lenz2020optimal}.
To propose an encoding algorithm to these codes, we need to propose encoding algorithms to RLL-SIDC codes and BSIDC codes.
Note that Saeki and Nozaki \cite{saeki2019} provided an encoding algorithm for the shifted VT codes.

The purpose of this research is to propose an encoding algorithm for an RLL-SIDC code.
To propose an efficient encoding algorithm for the RLL-SIDC code,
one might think that
we should modify the encoding algorithm of the binary VT code.
However,
in the systematic encoding algorithm for binary VT codes \cite{abdel1998systematic},
the parity part is at the positions of 2 powers.
The parity part must satisfy the run-length and congruence constraints.
Since it is scattered at the positions of 2 powers,
it is difficult to satisfy these two constraints at the same time.
Therefore, we need to consider another type of SIDC code.

Firstly, we construct a systematic-like encodable SIDC code,
which has a mechanism to limit the maximum run-length of the codeword.
The parity part of the proposed code is consecutive on the front part
and
consists of two type symbols,
namely, symbols to limit the maximum run-length of codewords
and symbols to satisfy a constraint defined by a linear congruence.

Secondly, we propose an encoding algorithm for the RLL-SIDC code.
It is a variation of modified concatenation \cite{mansuripur1991enumerative,immink1997practical} in constrained coding.
It works in the following procedure;
(i) The message is converted into a codeword in the $(0,r-1)$-constraint code by
Wijngaarden and Immink's algorithm \cite[Method C]{WijngaardenAndImmink2010code} (WI algorithm);
(ii) The $(0,r-1)$-constraint codeword is transformed into an $r$-RLL sequence
by the non-return-to-zero inverted (NRZI);
(iii) The encoding algorithm embeds this $r$-RLL sequence into
the message part of the SIDC code and computes the parity part.
Moreover, we clarify the parameters of the constructed code,
such that the proposed encoding algorithm works properly.

The remaining of the paper is organized as follows.
Section \ref{sec:pre} gives the notations used throughout the paper
and introduces existing algorithms.
Section \ref{sec:rll_encoder} shows that the RLL sequence encoder by the WI algorithm and the NRZI is better than the one by Schoeny et al.\ \cite[Appendix B]{schoeny2017codes}.
Section \ref{sec:method} constructs an SIDC code
and proposes its encoding algorithm.
Moreover, we prove this encoding algorithm outputs a maximum run-length limited sequence.
Furthermore, we compare the redundancy of the proposed encoding algorithm and the lower bound of the redundancy of the optimal RLL-SIDC code.

\section{Preliminaries} \label{sec:pre}
This section gives notations used throughout the paper.
This section also introduces existing algorithms, namely the WI algorithm and the NRZI, for constructing an encoding algorithm of the RLL codes.

\subsection{Notation}
Let $\mathbb{Z}$, $\mathbb{Z}^+$ be the set of integers and positive integers, respectively.
Let $[a,b]$ be the set of integers between $a$ and $b$,
i.e., $[a,b] := \{ i \in \mathbb{Z} \mid a \leq i \leq b \}$.
For $Z \subseteq \mathbb{Z}$,
denote its minimum and maximum, by $\underline{Z}$ and $\overline{Z}$, respectively.
For example,
if $Z = [a,b]$, $\underline{Z} = a$ and $\overline{Z} = b$ hold.
For $a,b \in \mathbb{Z}$ and $n \in \mathbb{Z}^+$,
denote $a \equiv b \pmod{n}$ if $(a-b)$ divides $n$.
Denote the exclusive OR (XOR), by $\oplus$.

Every positive integer $x \in [1,2^{k}-1]$ is
represented by $\sum_{i=0}^{k-1}g_i \cdot 2^i$ ($g_i \in \{ 0,1 \} $).
For a fixed $k \in \mathbb{Z}^+$,
we define $\mathrm{Le}_k : [0, 2^k - 1] \rightarrow \{ 0,1 \}^k$
as $\mathrm{Le}_k(x) = g_0 g_1 g_2 \cdots g_{k-1}$.
This mapping is called \textit{little-endian} of integer.
For example,
$\mathrm{Le}_3(4) = \mathtt{001}$
and
$\mathrm{Le}_4(11) = \mathtt{1101}$.

Denote concatenation of sequences $\boldsymbol{x}$ and $\boldsymbol{y}$, by $\boldsymbol{x}\boldsymbol{y}$.
Let $\lambda$ be the null string.
For sequence $\boldsymbol{x}$ and $i\in \mathbb{Z}^{+}$, recursively define $\boldsymbol{x}^i = \boldsymbol{x} \boldsymbol{x}^{i-1}$, where $\boldsymbol{x}^{0} = \lambda$.
The consecutive subsequence $(x_s, x_{s+1},\cdots ,x_t)$ for sequence $\boldsymbol{x}=(x_1, x_2, \cdots ,x_n)$ ($1 \leq s < t \leq n$)
is denoted by $\boldsymbol{x}_{[s,t]}$.
For $1 < i < j < n$,
the subsequence $\boldsymbol{x}_{[i,j]}$
is a run of length $r = j-i+1$
if $x_{i-1} \not = x_i = x_{i+1} = \cdots = x_{j} \not = x_{j+1}$.
As exceptions, $x_1$ is the start of a run and $x_n$ is the end of a run.
The length of the longest run in a sequence is called the \textit{maximum run-length}.
Let $S_{n,r}$ be
the set of sequences $\boldsymbol{x} \in \{ \mathtt{0}, \mathtt{1} \}^n$
whose maximum run-length is smaller than or equal to $r$.
The code $S_{n,r}$ is called $r$-RLL code of length $n$.

\subsection{$r$-RLL codes}
A $(0,r-1)$-constraint code is a set of sequences that the run-length of zero symbols are less than $r$.
The WI algorithm is known as an encoding algorithm for the $(0,r-1)$-constraint code.
In addition, the NRZI converts a $(0,r-1)$-constraint codeword to an $r$-RLL codeword.

\subsubsection{WI algorithm}
Let $\mathcal{G}_{r-1}^{(k)}$ be the $(0,r-1)$-constraint code of length $k$.
The WI algorithm converts a binary sequence $\vec{u}$ of length $k-1$
into a codeword $\vec{x}$ in $\mathcal{G}_{r-1}^{(k)}$.
Here, $k$ satisfies $k \leq 2^{r}+r-5$.
Roughly speaking, the WI algorithm repeats the \textit{replacement step}, which removes the \textit{forbidden word} $\mathtt{0}^{r}\mathtt{1}$ and attaches the sequence representing the position of removed forbidden word, while forbidden word exists.
After that, the WI algorithm attaches the sequence representing the number $s$ of replacement.

To explain the details of the algorithm, we introduce several notations.
We denote substitution $i$ for $j$, by $j \leftarrow i$.
The mapping $\omega_{r-1} : \mathbb{Z}^+ \to \{ \mathtt{0},\mathtt{1} \}^{*}$
gives the sequence representing the number $s$ of replacement.
The output is defined as follows:
\begin{align*}
  \omega_{r-1} (s)
  :=
  \mathtt{0}^{s - rv}(\mathtt{1}^{r-1} \mathtt{0})^v,
\end{align*}
where $v := \lfloor s/(k + 1) \rfloor$
and
$\lfloor \cdot \rfloor$ stands for the floor function.
For instance, the output of $\omega_4 (x)$ for $x \in [0,7]$ is summarized in Table \ref{tab:f_omega}.

\begin{table}[t]
  \centering
  \caption{Outputs of mapping $\omega_4 (x)$ for $x \in [0,7]$}
  \label{tab:f_omega}
  \scalebox{1.0}[1.0]{
    \begin{tabular}{|c|l||c|l|}
      \hline
      $x$ & \multicolumn{1}{c||}{$\omega_4 (x)$} & $x$ & \multicolumn{1}{c|}{$\omega_4 (x)$} \\ \hline
      0 & $\lambda$        & 4 & $\mathtt{0000}$ \\
      1 & $\mathtt{0}$     & 5 & $\mathtt{11110}$ \\
      2 & $\mathtt{00}$    & 6 & $\mathtt{011110}$ \\
      3 & $\mathtt{000}$   & 7 & $\mathtt{0011110}$ \\
      \hline
    \end{tabular}
  }
\end{table}

The details of the algorithm are described as follows:
\begin{enumerate}
  \item Set $s \leftarrow 0$, $\boldsymbol{v}_0 \leftarrow \boldsymbol{u}$.
  \item If $\vec{v}_s 1 \in \mathcal{G}_{(r-1)}^{k-s}$, go to Step 5.
  \item
    Search forbidden words $\mathtt{0}^{r}\mathtt{1}$ from the begging of $\boldsymbol{v}_s \mathtt{1}$.
    Set $p_s$ as the position where starts the forbidden word in the sequence $\boldsymbol{v}_s \mathtt{1}$.

  \item If $p_s < (k-s)-r$, remove $\mathtt{0}^{r}\mathtt{1}$ at $p_s$ from $\boldsymbol{v}_s$ and set $\boldsymbol{v}_{s+1} \leftarrow \boldsymbol{v}_s (\mathrm{Le}_{r-1}(p_s+3))$.
  If $p_s = (k-s)-r$, i.e., if it points to the end of $\vec{v}_{s}\mathtt{1}$, remove $\mathtt{0}^{r}$ from $\boldsymbol{v}_s$ and set $\boldsymbol{v}_{s+1} \leftarrow \boldsymbol{v}_s (\mathtt{1}\mathtt{0}^{r-2})$, $s \leftarrow s+1$.
  Return to Step 2.
  \item Set $\boldsymbol{w} \leftarrow \omega_k(s)$.
  Output $\vec{x} = \boldsymbol{v}_s \mathtt{1} \boldsymbol{w}$.
\end{enumerate}

\begin{ex.}\upshape
  \label{ex:WI}
  For
  $\vec{u} = (\mathtt{100000101000000100100000})$ and $r = 3$,
  the encoding process is as follows.
  \begin{enumerate}
    \item Set message $\vec{u}$ to $\vec{v}_0$
    and append $\mathtt{1}$.
    \begin{align*}
      \vec{v}_0 \mathtt{1}
      =
      (\mathtt{100000101000000100100000} \mathtt{1}).
    \end{align*}
    \item Repeat the conversion so that the run of zero symbols is $r-1$ or less.
    Table \ref{tab:WI} displays this operation.
    As a result, we get
    \begin{align*}
      \vec{v}_3 \mathtt{1}
      =
      (\mathtt{1010001001010100010001}).
    \end{align*}
    \item Set $\vec{w} \leftarrow \mathtt{000}$ and
    output $\vec{v}_3 \mathtt{1} \vec{w}$ as the codeword $\vec{x}$:
    \begin{align*}
      \vec{x}
      =
      (\mathtt{1010001000101000011011000}).
    \end{align*}
  \end{enumerate}

  \begin{table}[t]
  \centering
  \caption{Conversion operation for Example \ref{ex:WI}}
  \label{tab:WI}
  \scalebox{0.85}[0.85]{
    \begin{tabular}{|c|l|c|l|}
      \hline
      $s$ & \multicolumn{1}{c|}{$\vec{v}_s \boxed{\mathtt{1}} $} & $p_s$ & \multicolumn{1}{c|}{$(\mathrm{Le}_{r-1}(p_s+3))$} \\ \hline
      0   & $\mathtt{1\textcolor{red}{\underline{000001}}01000000100100000} \boxed{\mathtt{1}}$                               & 1 & $\mathtt{00101}$ \\
      1   & $\mathtt{1010\textcolor{red}{\underline{000001}}00100000 \textcolor{blue}{\underline{00101}}} \boxed{\mathtt{1}}$ & 5 & $\mathtt{01000}$ \\
      2   & $\mathtt{101000100\textcolor{red}{\underline{000001}} 01\textcolor{blue}{\underline{01000}}} \boxed{\mathtt{1}}$  & 10& $\mathtt{01101}$  \\
      3   & $\mathtt{1010001000101000\textcolor{blue}{\underline{01101}}} \boxed{\mathtt{1}}$  & - & \multicolumn{1}{c|}{-}  \\
      \hline
    \end{tabular}
  }
\end{table}
\end{ex.}

The decoding algorithm is described in \cite{WijngaardenAndImmink2010code}.

\subsubsection{NRZI} \label{sect:NRZI}
The NRZI converts a $(0,r - 1)$-constraint word $\boldsymbol{x}$
of length $k-1$ into $\boldsymbol{y}\in S_{k,r}$.
The encoding algorithm sets $y_{1} = x_{1}$
and
$y_i = y_{i+1} \oplus x_i$ for $i \geq 2$.

The decoder of the NRZI converts $\boldsymbol{y}\in S_{k,r}$ into $(0,r-1)$-constraint word $\boldsymbol{x}$ of length $k-1$.
The decoding algorithm sets $x_1=y_1$
and
$x_i = y_{i-1} \oplus y_i$ for $i \geq 2$.

\begin{ex.}\upshape
  For the following input $\vec{x}$, 
  the output $\vec{y}$ of the NRZI is
  \begin{align*}
    \vec{x} &= (\mathtt{1010001000101000011011000}), \\
    \vec{y} &= (\mathtt{1100001111001111101101111}).
  \end{align*}
\end{ex.}

\section{Comparison of $r$-RLL sequence encoders} \label{sec:rll_encoder}
In this section, we will show that the $r$-RLL sequence encoder by the WI algorithm and the NRZI
is better than the one by Schoeny et al.\ \cite[Appendix B]{schoeny2017codes}
from the relation between the code length $n$ and the maximum run-length $r$.

We give some lemmas to show that.

\begin{le.}\upshape
  For a fixed maximum run length $r$, 
  the code length $n$ of the $r$-RLL sequence by \cite[Appendix B]{schoeny2017codes} satisfies
  \begin{align}
    \label{eq:n_r_sch}
    n \leq 2^{r-3} + 1 =: G_r.
  \end{align}
\end{le.}

\begin{IEEEproof}
  In the method by \cite[Appendix B]{schoeny2017codes},
  the maximum run-length $r$ is satisfying as follows:
  \begin{align*}
    r = \lceil \log_2 (n-1) \rceil + 3.
  \end{align*}
  From this, we get Eq.\ \eqref{eq:n_r_sch}.
\end{IEEEproof}

\begin{le.}\upshape
  For a fixed maximum run length $r$, 
  the code length $n$ of the $r$-RLL sequence by the WI algorithm and the NRZI satisfies
  \begin{align}
    \label{eq:n_r_wi}
    n \leq 2^r + r - 5 =: H_r.
  \end{align}
\end{le.}

\begin{IEEEproof}
  In the method by the WI algorithm and the NRZI,
  code length $n$ is satisfying $n \leq 2^{r+1} + r - 4$.
  Moreover, from Sect.\ \ref{sect:NRZI}, 
  the maximum run-length becomes just $1$ longer.
  Combining these, we obtain Eq.\ \eqref{eq:n_r_wi}.
\end{IEEEproof}
Lemmas above lead the following remark.
\begin{re.}\upshape
  For $r \geq 3$,
  \begin{align*}
    H_r - G_r
    =
    7 \cdot 2^{r-3} + r - 6
    >
    0,
  \end{align*}
  holds.
  Hence, for a fixed maximum run-length $r$,
  when we use the method by the WI algorithm and the NRZI,
  the code length $n$ becomes longer.
  In other words,
  for a fixed code length $n$,
  to use the method by the WI algorithm and the NRZI,
  we can make the maximum run-length $r$ smaller.
\end{re.}
From this remark, in this paper, we use the method by the WI algorithm and the NRZI as an $r$-RLL sequence encoder.

\section{Efficient Encodable RLL-SIDC Code}
\label{sec:method}
In this section,
we construct an efficient encodable code correcting an insertion or deletion error
and propose its encoding algorithm.
Moreover,
we show that the outputs of this encoding algorithm are in the $r$-RLL code.
Furthermore, we compare the redundancy of the proposed encoding algorithm and the lower bound of the redundancy of the optimal RLL-SIDC code.

\subsection{Code construction}
\begin{def.}\upshape
  \label{def:C}
  Consider a sequence $ \boldsymbol{z} := (z_1, z_2, \cdots ,z_n) \in \{ \mathtt{0}, \mathtt{1} \}^n $.
  Suppose $\hat{r} \geq 4$ and $d_{\hat{r}} \in [2^{\hat{r}-2}+1,2^{\hat{r}-1}-1]$.
  For fixed $n, \hat{r}$, and $d_{\hat{r}}$,
  we define integer sequence $\{ a[n,\hat{r},d_{\hat{r}}]_i \}_{i=1}^{n+1}$ with length $n+1$ as follows:
  \begin{align}
    \label{eq:a}
    a[n,\hat{r},d_{\hat{r}}]_i :=
    \begin{cases}
      2^{i-1} & (1 \leq i < \hat{r}) \\
      d_{\hat{r}} & (i = \hat{r}) \\
      2^{i-2} & (i = \hat{r}+1, \hat{r}+2) \\
      2^{\hat{r}} - \hat{r} - 2 + i & (i \in [\hat{r} + 3, n+1]).
    \end{cases}
  \end{align}
  To simplify the notation, denote $a[n,\hat{r},d_{\hat{r}}]_i$, by $a_i$.
  Mapping $\mu : \{ \mathtt{0}, \mathtt{1} \}^n \rightarrow \mathbb{Z} $ is defined as follows:
  \begin{align*}
    \mu(\boldsymbol{z})
    :=
    \sum_{i=1}^{n}a[n,\hat{r},d_{\hat{r}}]_i z_i.
  \end{align*}
  For $b \in [0,a_{n+1}-1]$, we define a code
  \begin{align*}
    \boldsymbol{C}_b(n,\hat{r},d_{\hat{r}})
    := \{ \boldsymbol{z} \in \{ \mathtt{0}, \mathtt{1} \}^n : \mu(\boldsymbol{z}) \equiv b \pmod{a_{n+1}} \}.
  \end{align*}
\end{def.}

Note that the integer sequence $\{ a[n,\hat{r},d_{\hat{r}}]_i \}_{i=1}^{n+1}$ is positive monotonically increasing.
We give some examples of integer sequence $\{ a[n,\hat{r},d_{\hat{r}}]_i \}_{i=1}^{n+1}$.

\begin{ex.}\upshape
  Fix $d_4 \in [5,7]$ and $d_5 \in [9,15]$.
  The integer sequences $a[21,4,d_4]_i$ and $a[38,5,d_5]_i$ are
  \begin{align*}
    &\{ a[21,4,d_4]_i \}_{i=1}^{22} \\ &\quad= (1,2,4,d_4,8,16,17,18, \cdots ,30,31,32), \\
    &\{ a[38,5,d_5]_i \}_{i=1}^{39} \\ &\quad= (1,2,4,8,d_5,16,32,33,34, \cdots ,62,63,64).
  \end{align*}
\end{ex.}

The following theorem shows error correcting capability of $\boldsymbol{C}_b(n,\hat{r},d_{\hat{r}})$.
\begin{th.}\upshape
  \label{th:mycode}
  For any $n\in \mathbb{Z}^+$, $\hat{r}\ge 4$, $d_{\hat{r}} \in [2^{\hat{r}-2}+1,2^{\hat{r}-1}-1]$, and $b\in[0,a_{n+1}-1]$,
  $\boldsymbol{C}_b(n,\hat{r},d_{\hat{r}})$ is an SIDC code.
\end{th.}
\begin{IEEEproof}
Recall that $\{ a[n,\hat{r},d_{\hat{r}}]_i \}_{i=1}^{n+1}$ is a positive monotonically sequence.
Hence, $\{ a[n,\hat{r},d_{\hat{r}}]_i \}_{i=1}^{n+1}$ is a monotonically increasing code \cite{hagiwara2016ordered}.
Since monotonically increasing codes are SIDC codes \cite{hagiwara2016ordered},
$\boldsymbol{C}_b(n,\hat{r},d_{\hat{r}})$ is an SIDC code.
\end{IEEEproof}

\subsection{Encoding algorithm} \label{sec:encoding}

In this section,
we propose a systematic encoding algorithm for $\boldsymbol{C}_b(n,\hat{r},d_{\hat{r}})$.

\subsubsection{Overview}
\begin{figure}[t]
  \begin{picture}(250,160)
    \put(10,20){\includegraphics[width=170pt,height=118pt]{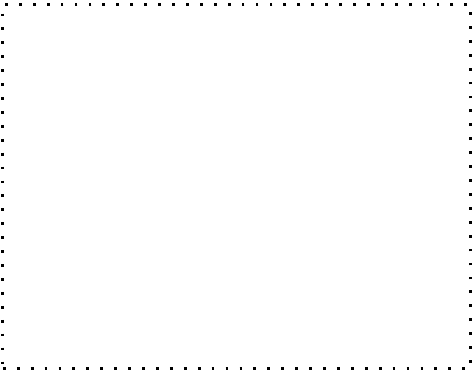}}

    \put(182,108){$\leftarrow$ Proposed}
    \put(196,96){encoding}
    \put(196,84){algorithm}

    \put(31,147){$\boldsymbol{u} \in \{ \mathtt{0,1} \}^{k-1}$}
    \put(30,131){{\Large$\Downarrow$}}
    \put(20,108){\framebox(65,15){WI algorithm}}
    \put(45,94){$\boldsymbol{x} \in \mathcal{G}_{r-1}^{(k)}$ : $(0,r-1)$-constraint}
    \put(155,82){code}
    \put(30,93){{\Large$\Downarrow$}}
    \put(20,70){\framebox(30,15){NRZI}}
    \put(45,56){$\boldsymbol{y} \in S_{k,r}$ : $r$-RLL code}
    \put(30,55){{\Large$\Downarrow$}}
    \put(20,32){\framebox(70,15){Algorithm \ref{alg:enc}}}
    \put(30,17){{\Large$\Downarrow$}}
    \put(31,6){$\boldsymbol{z} \in \boldsymbol{C}_b(n,\hat{r},d_{\hat{r}}) \cap S_{n,r}$ : RLL-SIDC code}
  \end{picture}
  \caption{Flow of encoding}
  \label{fig:ov}
\end{figure}
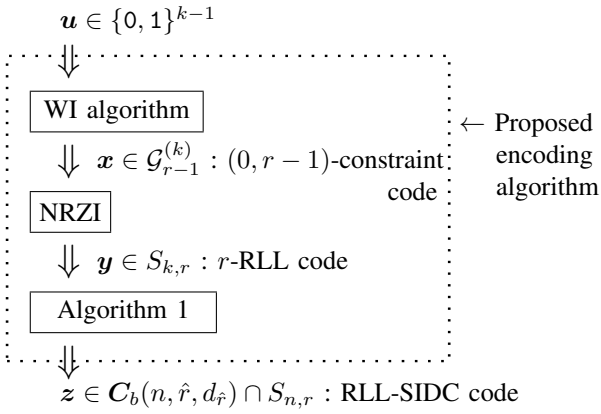

This algorithm converts a binary sequence $\vec{u} \in \{0,1\}^{k-1}$
into the codeword
$\vec{z} \in \boldsymbol{C}_b(n,\hat{r},d_{\hat{r}}) \cap S_{n,r}$.
Figure \ref{fig:ov} depicts the flow of the message.
The original message $\boldsymbol{u} \in \{ \mathtt{0,1} \}^{k-1}$ is converted into $\boldsymbol{y} \in S_{k,r}$
by the WI algorithm and the NRZI.
Algorithm \ref{alg:enc},
whose details will be shown in Section \ref{sssec:Algo},
converts $\vec{y} \in S_{k,r}$ into $\vec{z} \in \boldsymbol{C}_b(n,\hat{r},d_{\hat{r}}) \cap S_{n,r}$.

The output is represented by $\boldsymbol{z} = \boldsymbol{py}$,
where $\boldsymbol{p}$ and $\boldsymbol{y}$ stand for
the parity and the message parts, respectively.
The symbols of the parity part are divided into two types.
One of them limits the run-length of a codeword.
The other type is used for satisfying $\mu(\boldsymbol{z}) \equiv b \pmod{a_{n+1}}$.
Algorithm \ref{alg:enc} is divided into three stages.
At the first stage, $\boldsymbol{y} \in S_{k,r}$ is embedded in the message part.
At the second stage, two specific symbols are decided to limit the run-length of a codeword.
At the last stage, other symbols are computed as satisfying $\mu(\boldsymbol{z}) \equiv b \pmod{a_{n+1}}$.

\subsubsection{Encoding algorithm \label{sssec:Algo}}
Fix the length of message part $k \geq 7$.
We decide $\hat{r}$ as follows:
\begin{align}
  \label{eq:hat_r}
  \hat{r} = \lceil \log_{2} (k+2) \rceil,
\end{align}
where $\lceil \cdot \rceil$ stands for the ceiling function.
Let $m$ be the length of the parity part
$\boldsymbol{p}:=(p_1, p_2, \cdots ,p_{\hat{r}} ,p_{\hat{r}+1} ,p_{\hat{r}+2} ,p_m) \in \{ \mathtt{0}, \mathtt{1} \}^{m}$
and $m$ satisfies
\begin{align}
  \label{eq:m}
  m = \hat{r} + 3.
\end{align}
The code length $n$ satisfies 
\begin{align}
  \label{eq:n}
  n = m + k.
\end{align}
This algorithm requires parameters
$r \geq \hat{r}$,
$d_{\hat{r}} \in [2^{\hat{r}-2}+1,2^{\hat{r}-1}-1]$,
and $b \in [0,a_{n+1}-1]$,
as input,
where $a_{n+1} = 2^{\hat{r}} + k + 2$ from Eq.\ \eqref{eq:a}.

The input of the algorithm is $\vec{y} \in S_{k,r}$.
The output of the algorithm is $\vec{z} \in \boldsymbol{C}_b(n,\hat{r},d_{\hat{r}}) \cap S_{n,r}$,
where $\vec{z} = \vec{py}$.
Symbols $p_{\hat{r}}, p_m$ are used for limiting the run-length of the codeword.
More precisely,
symbol $p_m = y_1 \oplus \mathtt{1}$ separates the run of the parity part and the message part.
Symbol $p_{\hat{r}}$ limits the run-length of the parity part.
Provisionally, set $p_{\hat{r}} = \mathtt{0}$.
If the run-length of the parity part exceeds the limit,
the symbol is changed $p_{\hat{r}}= \mathtt{1}$.
We will show this ad hoc method always limits the run length of parity part in Sect.\ \ref{sec:RLL}.
The other symbols of the parity part are computed for satisfying $\mu(\boldsymbol{z}) \equiv b \pmod{a_{n+1}}$.
Define
\begin{align*}
  \boldsymbol{q}
  :=
  ( p_1, p_2, \cdots ,p_{\hat{r}-1},p_{\hat{r}+1},p_{\hat{r}+2})
  \in
  \{ \mathtt{0}, \mathtt{1} \}^{\hat{r}+1}.
\end{align*}
Table \ref{tab:a2y} shows $a_i$ and $p_i$ for $i \in [1,m]$.
From this table,
we see that $a_i$ is a power of $2$ for $i \in [1,m-1] \setminus \{ \hat{r} \}$.
Hence, $\mu(\boldsymbol{py}) \equiv b \pmod{a_{n+1}}$ is equivalent to
\begin{align*}
  &\sum_{i=0}^{\hat{r}} 2^i q_i \\
  \equiv&~
  b - d_{\hat{r}} p_{\hat{r}} - a_m p_m - \sum_{j=1}^{k}a_{j+m} y_j \pmod{a_{n+1}}.
\end{align*}
From this, we determine $\boldsymbol{q}$ uniquely once $p_{\hat{r}},p_m$ and $\boldsymbol{y}$ are decided.
Algorithm \ref{alg:enc} summarizes the procedure above.

\begin{table}[t]
  \centering
  \caption{The value $a_i$ and $p_i$ for $i \in [1,m]$}
  \label{tab:a2y}
  \scalebox{.95}[.95]{
  \begin{tabular}{|c||cccccccc|}
    \hline
    $a_i$ & $2^0$ & $2^1$ & $\cdots$ & $2^{\hat{r}-2}$ & $d_{\hat{r}}$ & $2^{\hat{r}-1}$ & $2^{\hat{r}}$ & $2^{\hat{r}}+1$ \\
    $p_i$ & $p_1$ & $p_2$ & $\cdots$ & $p_{\hat{r}-1}$ & $p_{\hat{r}}$ & $p_{\hat{r}+1}$ & $p_{\hat{r}+2}$ & $p_m$ \\ \hline
  \end{tabular}
  }
\end{table}

\begin{algorithm}[t]
\caption{Conversion RLL sequence into RLL-SIDC codeword}
\label{alg:enc}
\begin{algorithmic}[1]
\REQUIRE Parameter $r$,$d_{\hat{r}}$,$b$, Sequence $\boldsymbol{y} \in S_{k,r}$
\ENSURE Sequence $\boldsymbol{z} \in \boldsymbol{C}_b(n,\hat{r},d_{\hat{r}}) \cap S_{n,r}$
\STATE Set $p_m \leftarrow y_1 \oplus \mathtt{1}$, \  $p_{\hat{r}} \leftarrow \mathtt{0}$
\STATE Calculate $\boldsymbol{q}$ satisfying $\mu(\boldsymbol{p} \boldsymbol{y}) \equiv b \pmod{a_{n+1}}$
\label{alg:z}
\IF{(length of run in  $ \boldsymbol{p}$) $> r$}
\label{alg:if}
\STATE Set $p_{\hat{r}} \leftarrow \mathtt{1}$
\STATE Calculate $\boldsymbol{q}$ satisfying $\mu(\boldsymbol{p} \boldsymbol{y}) \equiv b \pmod{a_{n+1}}$
\label{alg:o}
\ENDIF
\STATE Output $\boldsymbol{z} = \boldsymbol{py}$
\end{algorithmic}
\end{algorithm}

\begin{ex.}\upshape
  For $\boldsymbol{y}=(\mathtt{10100001000010})$,
  $\hat{r}=4,r=4,d_{\hat{r}}=6,b=31$,
  the process of the encoding algorithm is as follows:
  \begin{enumerate}
    \item Embed $\boldsymbol{y}$ into the message part,
    \begin{align*}
      \boldsymbol{py}=(p_1p_2p_3p_{\hat{r}}p_5p_6p_m\mathtt{10100001000010}).
    \end{align*}
    \item Set $p_m = \mathtt{0}$ and $p_{\hat{r}} = \mathtt{0}$,
    \begin{align*}
      \boldsymbol{py}=(p_1p_2p_3\mathtt{0}p_5p_6\mathtt{010100001000010}).
    \end{align*}
    \item Compute $\boldsymbol{q}$ as satisfy $\mu(\boldsymbol{z}) \equiv 31 \pmod{32}$,
    \begin{align*}
      \boldsymbol{py}=(\mathtt{010000010100001000010}).
    \end{align*}
    \item Since the maximum run-length of the parity part $\boldsymbol{p}=(\mathtt{0100000})$ exceed $r$, the encoding algorithm resets $p_{\hat{r}} = \mathtt{1}$,
    \begin{align*}
      \boldsymbol{py}=(p_1p_2p_3\mathtt{1}p_5p_6\mathtt{010100001000010}).
    \end{align*}
    \item Determine $\boldsymbol{q}$ to satisfy $\mu(\boldsymbol{py}) \equiv b \pmod{a_{n+1}}$ again.
    Thereafter, output the sequence $\boldsymbol{py}$ as codeword $\boldsymbol{z}$,
    \begin{align*}
      \boldsymbol{z}=(\mathtt{001111010100001000010}).
    \end{align*}
  \end{enumerate}

  Table \ref{tab:enc4} summarizes the change of the parity part.
  
  \begin{table}[t]
    \centering
    \caption{The change of the parity part for $\boldsymbol{y}=(10100001000010)$,$\hat{r}=4$,$r=4$,$d_{\hat{r}}=6$,$b=31$}
    \scalebox{.8}[.8]{
    \begin{tabular}{|c||ccccccc|}
      \hline
      $i$ & 1 & 2 & 3 & 4 & 5 & 6 & 7 \\
      $\boldsymbol{z}$ & $p_1$ & $p_2$ & $p_3$ & $p_{\hat{r}}$ & $p_5$ & $p_6$ & $p_m$  \\
      $a_i$ & 1 & 2 & 4 & 6 & 8 & 16 & 17  \\ \hline
      \textbf{Step 2} & $p_1$ & $p_2$ & $p_3$ & $\mathtt{0}$ & $p_5$ & $p_6$ & $\mathtt{0}$  \\
      \textbf{Step 3} & $\mathtt{0}$ & $\mathtt{1}$ & $\mathtt{0}$ & $\mathtt{0}$ & $\mathtt{0}$ & $\mathtt{0}$ & $\mathtt{0}$  \\
      \textbf{Step 4} & $p_1$ & $p_2$ & $p_3$ & $\mathtt{1}$ & $p_5$ & $p_6$ & $\mathtt{0}$  \\
      \textbf{Step 5} & $\mathtt{0}$ & $\mathtt{0}$ & $\mathtt{1}$ & $\mathtt{1}$ & $\mathtt{1}$ & $\mathtt{1}$ & $\mathtt{0}$  \\ \hline
    \end{tabular}
    }
    \label{tab:enc4}
  \end{table}
\end{ex.}

\subsection{Decoding algorithm}
Recall that $\boldsymbol{C}_b(n,\hat{r},d_{\hat{r}})$ is a monotonically increasing code.
Hence,
we get decoding algorithm $\boldsymbol{C}_b(n,\hat{r},d_{\hat{r}})$ by applying \cite{dec}.

\subsection{Run-length limited property and proof}
\label{sec:RLL}

Theorem \ref{th:RLL_under} shows that Algorithm \ref{alg:enc} limits the maximum run-length when $\hat{r} = r$.
\begin{th.}\upshape
  \label{th:RLL_under}
  Suppose input $\boldsymbol{y}$ is in $S_{k,r}$.
  If $r = \hat{r} \geq 4$ and $(k, r, d_{\hat{r}}) \not = (14,4,5)$,
  for all $b \in [0,a_{n+1}-1]$,
  the output $\boldsymbol{z}$ of Algorithm \ref{alg:enc} satisfies $\boldsymbol{z} \in S_{n,r}$.
\end{th.}
\begin{IEEEproof}
  For a given $\boldsymbol{y}$, let $\boldsymbol{z}^{(\mathtt{0})},\boldsymbol{z}^{(\mathtt{1})}$ be
  the sequences obtained by Step \ref{alg:z}, \ref{alg:o} of Algorithm \ref{alg:enc}, respectively.
  Hence,
  $\boldsymbol{z}^{(\mathtt{0})},\boldsymbol{z}^{(\mathtt{1})} \in \boldsymbol{C}_b(n,\hat{r},d_{\hat{r}})$
  and
  $\boldsymbol{z}_{[m+1,n]}^{(\mathtt{0})} = \boldsymbol{z}_{[m+1,n]}^{(\mathtt{1})} = \boldsymbol{y}$ hold.
  In addition,
  $z_{m}^{(\mathtt{0})} = z_{m}^{(\mathtt{1})}$,
  $z_{\hat{r}}^{(\mathtt{0})} = \mathtt{0},$ and $z_{\hat{r}}^{(\mathtt{1})} = \mathtt{1}$ also hold.

  Since $z_{m}^{(\mathtt{0})} = z_{m}^{(\mathtt{1})} \not = z_{m+1}^{(\mathtt{0})} = z_{m+1}^{(\mathtt{1})}$,
  the $m$-th and $(m+1)$-th symbols are belong to distinct runs.
  Moreover,
  since $\boldsymbol{y} \in S_{k,r}$,
  the maximum run-length of the message part is smaller than or equal to $r$.
  Hence, if run-length of the parity part is smaller than or equal to $r$,
  the maximum run-length of the codeword is also smaller than or equal to $r$.
  Thereby, we prove using contradiction to be the maximum run-length limited
  either
  $\boldsymbol{p}^{(\mathtt{0})} := \boldsymbol{z}_{[1,m]}^{(\mathtt{0})}$
  or
  $\boldsymbol{p}^{(\mathtt{1})} := \boldsymbol{z}_{[1,m]}^{(\mathtt{1})}$.

  Define mappings
  $\rho : \{ \mathtt{0}, \mathtt{1} \}^{m} \rightarrow \mathbb{Z}$
  and
  $\sigma : \{ \mathtt{0}, \mathtt{1} \}^{k} \rightarrow \mathbb{Z}$
  as follows:
  \begin{align*}
    \rho (\boldsymbol{p}) := \sum_{\substack{i=1 \\ i \not = \hat{r}}}^{m-1}a_ip_i,
    \qquad
    \sigma (\boldsymbol{y}) := \sum_{i=m+1}^{n}a_iy_{i-m}.
  \end{align*}
  Then, the mapping $\mu$ is rewritten by
  \begin{align}
    \label{eq:sum_all}
    \mu(\boldsymbol{z})
    =
    \rho(\boldsymbol{p}) + dz_{\hat{r}} + a_mz_m + \sigma(\boldsymbol{y}).
  \end{align}
  Since $\boldsymbol{z}^{(\mathtt{0})},\boldsymbol{z}^{(\mathtt{1})} \in \boldsymbol{C}_b(n,\hat{r},d_{\hat{r}})$,
  we get
  \begin{align*}
      &\mu(\boldsymbol{z}^{(\mathtt{0})}) \equiv b \pmod{a_{n+1}}, \\
      &\mu(\boldsymbol{z}^{(\mathtt{1})}) \equiv b \pmod{a_{n+1}}.
  \end{align*}
  This yields
  \begin{align*}
    \mu(\boldsymbol{z}^{(\mathtt{0})}) - \mu(\boldsymbol{z}^{(\mathtt{1})}) \equiv 0 \pmod{a_{n+1}}.
  \end{align*}
  Recall that
  $z_{\hat{r}}^{(\mathtt{0})} = \mathtt{0}$, $z_{\hat{r}}^{(\mathtt{1})} = \mathtt{1}$,
  $z_{m}^{(\mathtt{0})} = z_{m}^{(\mathtt{1})}$
  and
  $\boldsymbol{z}_{[m+1,n]}^{(\mathtt{0})} = \boldsymbol{z}_{[m+1,n]}^{(\mathtt{1})}$.
  Combining Eq.\ \eqref{eq:sum_all}, we have
  \begin{align*}
    \rho(\boldsymbol{p}^{(\mathtt{1})})+d_{\hat{r}}
    - \rho(\boldsymbol{p}^{(\mathtt{0})}) \equiv 0 \pmod{a_{n+1}}.
  \end{align*}
  We denote the left hand side of this congruence, by
  $A(\boldsymbol{p}^{(\mathtt{0})},\boldsymbol{p}^{(\mathtt{1})},d_{\hat{r}})$.
  The congruence above shows that there exists $l \in \mathbb{Z}$
  such that
  \begin{align}
    \label{eq:A01}
    A(\boldsymbol{p}^{(\mathtt{0})},\boldsymbol{p}^{(\mathtt{1})},d_{\hat{r}})
    =
    la_{n+1}.
  \end{align}

  Let us evaluate
  $A(\boldsymbol{p}^{(\mathtt{0})},\boldsymbol{p}^{(\mathtt{1})},d_{\hat{r}})$.
  We should consider the two cases, namely
  (i) $z_m^{(\mathtt{0})} = z_m^{(\mathtt{1})} = \mathtt{0}$
  and
  (ii) $z_m^{(\mathtt{0})} = z_m^{(\mathtt{1})} = \mathtt{1}$.
  We show the proof in the case of
  $z_m^{(\mathtt{0})} = z_m^{(\mathtt{1})} = \mathtt{0}$.
  The \textit{zero forbidden words} (resp. \textit{one forbidden words}) are
  parity parts containing run of $\mathtt{0}$ (resp. $\mathtt{1}$) of length at least $r$.
  We enumerate the zero and one forbidden words as follows:
  \begin{align*}
    \boldsymbol{f}_{\mathtt{0},1} &:= (\mathtt{0},\mathtt{0},\mathtt{0}^{\hat{r}-1},\mathtt{0},\mathtt{0}), \quad
    \boldsymbol{f}_{\mathtt{0},2} := (\mathtt{1},\mathtt{0},\mathtt{0}^{\hat{r}-1},\mathtt{0},\mathtt{0}), \\
    \boldsymbol{f}_{\mathtt{0},3} &:= (\mathtt{0},\mathtt{1},\mathtt{0}^{\hat{r}-1},\mathtt{0},\mathtt{0}), \quad
    \boldsymbol{f}_{\mathtt{0},4} := (\mathtt{1},\mathtt{1},\mathtt{0}^{\hat{r}-1},\mathtt{0},\mathtt{0}), \\
    \boldsymbol{f}_{\mathtt{0},5} &:= (\mathtt{0},\mathtt{0},\mathtt{0}^{\hat{r}-1},\mathtt{1},\mathtt{0}), \\
    \boldsymbol{f}_{\mathtt{1},1} &:= (\mathtt{1},\mathtt{1},\mathtt{1}^{\hat{r}-1},\mathtt{1},\mathtt{0}), \quad
    \boldsymbol{f}_{\mathtt{1},2} := (\mathtt{0},\mathtt{1},\mathtt{1}^{\hat{r}-1},\mathtt{1},\mathtt{0}), \\
    \boldsymbol{f}_{\mathtt{1},3} &:= (\mathtt{1},\mathtt{1},\mathtt{1}^{\hat{r}-1},\mathtt{0},\mathtt{0}).
  \end{align*}
  Table \ref{tab:0_seq} gives all the forbidden words and their mapping output $\rho(\boldsymbol{f}_{ij})$.
  From this table, we see that
  all the zero (resp. one) forbidden words satisfy
  $p_{\hat{r}} = \mathtt{0}$ (resp. $p_{\hat{r}} = \mathtt{1}$).
  Hence, if $\boldsymbol{z}^{(\mathtt{0})},\boldsymbol{z}^{(\mathtt{1})} \not \in S_{n,r}$,
  there exist $i,j$
  such that
  $\boldsymbol{p}^{(\mathtt{0})} = \boldsymbol{f}_{\mathtt{0},i}$
  and
  $\boldsymbol{p}^{(\mathtt{1})} = \boldsymbol{f}_{\mathtt{1},j}$.

  \begin{table}[t]
    \centering
    \caption{Forbidden words and mapping $\rho$}
    \label{tab:0_seq}
    \scalebox{.83}[.83]{
    \begin{tabular}{|c||ccccccc|c|}
      \hline
       & $p_1$ & $p_2$  &  $\cdots$  &  $p_{\hat{r}}$  & $p_{\hat{r}+1}$  &  $p_{\hat{r}+2}$ &  $p_m$ & $\rho(\boldsymbol{f}_{ij})$ \\
      $i$ & 1 &  2 & $\cdots$ & $\hat{r}$ & $\hat{r}+1$ & $\hat{r}+2$ & $m$ & \\ \hline
      $a_i$ & $2^0$ &  $2^1$  & $\cdots$ & $d_{\hat{r}}$ & $2^{\hat{r}-1}$ & $2^{\hat{r}}$ & $2^{\hat{r}}+1$  & \\ \hline
      $\boldsymbol{f}_{\mathtt{0},1}$ &  $\mathtt{0}$  &  $\mathtt{0}$  &  $\mathtt{0}^{\hat{r}-3}$  &  $\mathtt{0}$  &  $\mathtt{0}$  &  $\mathtt{0}$  &  $\mathtt{0}$  &  0\\
      $\boldsymbol{f}_{\mathtt{0},2}$ &  $\mathtt{1}$  &  $\mathtt{0}$  &  $\mathtt{0}^{\hat{r}-3}$  &  $\mathtt{0}$  &  $\mathtt{0}$  &  $\mathtt{0}$  &  $\mathtt{0}$  &  1\\
      $\boldsymbol{f}_{\mathtt{0},3}$ &  $\mathtt{0}$  &  $\mathtt{1}$  &  $\mathtt{0}^{\hat{r}-3}$  &  $\mathtt{0}$  &  $\mathtt{0}$  &  $\mathtt{0}$  &  $\mathtt{0}$  &  2\\
      $\boldsymbol{f}_{\mathtt{0},4}$ &  $\mathtt{1}$  &  $\mathtt{1}$  &  $\mathtt{0}^{\hat{r}-3}$  &  $\mathtt{0}$  &  $\mathtt{0}$  &  $\mathtt{0}$  &  $\mathtt{0}$  &  3\\
      $\boldsymbol{f}_{\mathtt{0},5}$ &  $\mathtt{0}$  &  $\mathtt{0}$  &  $\mathtt{0}^{\hat{r}-3}$  &  $\mathtt{0}$  &  $\mathtt{0}$  &  $\mathtt{1}$  &  $\mathtt{0}$  &  $2^{\hat{r}}$ \\ \hline
      $\boldsymbol{f}_{\mathtt{1},1}$ &  $\mathtt{1}$  &  $\mathtt{1}$  &  $\mathtt{1}^{\hat{r}-3}$  &  $\mathtt{1}$  &  $\mathtt{1}$  &  $\mathtt{1}$  &  $\mathtt{0}$  & $2^{\hat{r}+1}-1$ \\
      $\boldsymbol{f}_{\mathtt{1},2}$ &  $\mathtt{0}$  &  $\mathtt{1}$  &  $\mathtt{1}^{\hat{r}-3}$  &  $\mathtt{1}$  &  $\mathtt{1}$  &  $\mathtt{1}$  &  $\mathtt{0}$  & $2^{\hat{r}+1}-2$ \\
      $\boldsymbol{f}_{\mathtt{1},3}$ &  $\mathtt{1}$  &  $\mathtt{1}$  &  $\mathtt{1}^{\hat{r}-3}$  &  $\mathtt{1}$  &  $\mathtt{1}$  &  $\mathtt{0}$  &  $\mathtt{0}$  & $2^{\hat{r}}-1$ \\ \hline
    \end{tabular}
    }
  \end{table}

  \begin{table}[t]
    \centering
    \caption{Values of $A(\boldsymbol{f}_{\mathtt{0},i},\boldsymbol{f}_{\mathtt{1},j},d_{\hat{r}})$ for $z_m^{(\mathtt{0})} = z_m^{(\mathtt{0})} =  \mathtt{0}$}
    \label{tab:A0}
    \begin{tabular}{|c||c|cc|}
      \hline
      & $\boldsymbol{f}_{\mathtt{1},3}$ & $\boldsymbol{f}_{\mathtt{1},2}$ & $\boldsymbol{f}_{\mathtt{1},1}$ \\ \hline \hline
      $\boldsymbol{f}_{\mathtt{0},5}$ & $d_{\hat{r}}-1$ & $d_{\hat{r}}+2^{\hat{r}}-2$ & $d_{\hat{r}}+2^{\hat{r}}-1$ \\ \hline
      $\boldsymbol{f}_{\mathtt{0},4}$ & $d_{\hat{r}}+2^{\hat{r}}-4$ & $d_{\hat{r}}+2^{\hat{r}+1}-5$ & $d_{\hat{r}}+2^{\hat{r}+1}-4$ \\
      $\boldsymbol{f}_{\mathtt{0},3}$ & $d_{\hat{r}}+2^{\hat{r}}-3$ & $d_{\hat{r}}+2^{\hat{r}+1}-4$ & $d_{\hat{r}}+2^{\hat{r}+1}-3$ \\
      $\boldsymbol{f}_{\mathtt{0},2}$ & $d_{\hat{r}}+2^{\hat{r}}-2$ & $d_{\hat{r}}+2^{\hat{r}+1}-3$ & $d_{\hat{r}}+2^{\hat{r}+1}-2$ \\
      $\boldsymbol{f}_{\mathtt{0},1}$ & $d_{\hat{r}}+2^{\hat{r}}-1$ & $d_{\hat{r}}+2^{\hat{r}+1}-2$ & $d_{\hat{r}}+2^{\hat{r}+1}-1$ \\ \hline
    \end{tabular}
  \end{table}

  Hypothesize
  $\boldsymbol{p}^{(\mathtt{0})} = \boldsymbol{f}_{\mathtt{0},i}$
  and
  $\boldsymbol{p}^{(\mathtt{1})} = \boldsymbol{f}_{\mathtt{1},j}$
  ($i \in [1,5], j \in [1,3]$).
  Then,
  $A(\boldsymbol{f}_{\mathtt{0},i},\boldsymbol{f}_{\mathtt{1},j},d_{\hat{r}})$
  takes the value given in Table \ref{tab:A0}.
  Define
  \begin{align*}
    &B_1(d_{\hat{r}}) := \{ d_{\hat{r}}-1 \}, \\
    &B_2(d_{\hat{r}}) := [d_{\hat{r}}+2^{\hat{r}}-4,d_{\hat{r}}+2^{\hat{r}}-1], \\
    &B_3(d_{\hat{r}}) := [d_{\hat{r}}+2^{\hat{r}+1}-5,d_{\hat{r}}+2^{\hat{r}+1}-1].
  \end{align*}
  Then, from Table \ref{tab:A0},
  $A(\boldsymbol{f}_{\mathtt{0},i},\boldsymbol{f}_{\mathtt{1},j},d_{\hat{r}})$ takes
  value in $B_1(d_{\hat{r}}) \cup B_2(d_{\hat{r}}) \cup B_3(d_{\hat{r}})$.
  Recall $d_{\hat{r}} \in [2^{\hat{r}-2}+1,2^{\hat{r}-1}-1] := K$.
  Define $C_i := \cup_{d_{\hat{r}} \in K} B_i(d_{\hat{r}})$ for $i=1,2,3$.
  Then, we get
  \begin{align}
    &C_1 = [2^{\hat{r}-2},2^{\hat{r}-1}-2],     \label{eq:c1} \\
    &C_2 = [5 \cdot 2^{\hat{r}-2}-3,3 \cdot 2^{\hat{r}-1}-2],  \label{eq:c2} \\
    &C_3 = [9 \cdot 2^{\hat{r}-2}-4,5 \cdot 2^{\hat{r}-1}-2].  \label{eq:c3}
  \end{align}
  Recall that $a_{n+1} = 2^{\hat{r}} + k + 2$.
  Since $\hat{r} = \lceil \log_{2} (k+2) \rceil$ holds,
  we get $k \in [2^{\hat{r}-1}-1,2^{\hat{r}}-2]$ for a given $\hat{r}$.
  Define
  \begin{align}
    \label{eq:d}
    D := [3 \cdot 2^{\hat{r}-1} +1,2^{\hat{r}+1}].
  \end{align}
  Then, $a_{n+1}\in D$.

  \begin{figure}[t]
    \begin{picture}(230,47)
      \put(0,0){\includegraphics[width=230pt]{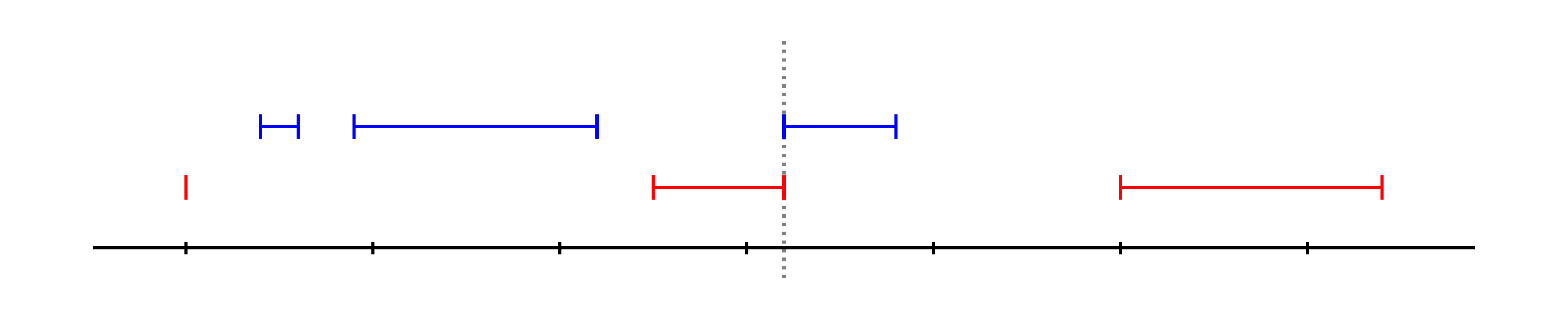}}
      \put(37,32){$C_1$}
      \put(64,32){$C_2$}
      \put(119,32){$C_3$}
      \put(25,22){$0$}
      \put(101.5,22){$D$}
      \put(178,22){$2D$}
    \end{picture}
    \caption{Intervals of $C_1$, $C_2$, $C_3$, $D$, and $2D$}
    \label{fig:r4}
  \end{figure}
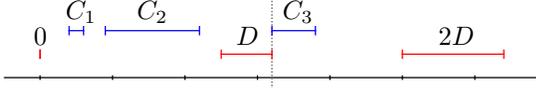
  
  For $l \in \mathbb{Z}$
  and
  $E \subseteq \mathbb{Z}$,
  we define $lE := \{ le : e \in E \}$.
  Now, we will give a necessary and sufficient condition for holding
  \begin{align}
    \label{eq:cd}
    \emptyset = (C_1 \cup C_2 \cup C_3) \cap \left(\bigcup_{l \in \mathbb{Z}}lD\right),
  \end{align}
  i.e., contradicting Eq.\ \eqref{eq:A01}.
  Figure \ref{fig:r4} depicts the intervals of
  $C_1$, $C_2$, $C_3$, $D$, and $2D$.
  From Eqs.\ \eqref{eq:c1}, \eqref{eq:c2}, \eqref{eq:c3}, and \eqref{eq:d},
  for $\hat{r} \geq 4$, we get
  \begin{align*}
    0 < \underline{C_1} < \overline{C_1} < \underline{C_2} < \overline{C_2} < \underline{D} < \overline{D} \overset{\text{(a)}}{\leq} \underline{C_3} < \overline{C_3} < \underline{2D}.
  \end{align*}
  Here equality (a) holds if and only if $\hat{r} = 4$.
  For $\hat{r} = 4$,
  $A(\boldsymbol{p}^{(\mathtt{0})}, \boldsymbol{p}^{(\mathtt{1})}, d_{\hat{r}})$
  achieves $\underline{C_3} = 32$ if and only if
  $\boldsymbol{p}^{(\mathtt{0})} = \boldsymbol{f}_{\mathtt{0},4}$,
  $\boldsymbol{p}^{(\mathtt{1})} = \boldsymbol{f}_{\mathtt{1},2}$,
  and
  $d_{\hat{r}} = 5$.
  Moreover,
  $a_{n+1}$ achieves $\overline{D} = 32$ if and only if
  $k = 14$
  for $\hat{r} = 4$.
  Therefore,
  a necessary and sufficient condition for holding Eq.\ \eqref{eq:cd} is
  \begin{align}
    \label{eq:krd}
    \hat{r} \geq 4 \quad \textrm{and}  \quad (k,r,d_{\hat{r}}) \not = (14,4,5).
  \end{align}
  By the proof by contradiction,
  we obtain that the output $\vec{Z}$ of Algorithm \ref{alg:enc} is
  in $S_{n,r}$ if Eq.\ \eqref{eq:krd} holds.

  Next, we consider the case of $z_m^{(\mathtt{0})} = z_m^{(\mathtt{1})} = \mathtt{1}$.
  Then the forbidden words are as follows:

  \begin{align*}
    \boldsymbol{f}_{\mathtt{1},4} &:= (\mathtt{1},\mathtt{1},\mathtt{1}^{\hat{r}-1},\mathtt{1},\mathtt{1}), \quad
    \boldsymbol{f}_{\mathtt{1},5} := (\mathtt{0},\mathtt{1},\mathtt{1}^{\hat{r}-1},\mathtt{1},\mathtt{1}), \\
    \boldsymbol{f}_{\mathtt{1},6} &:= (\mathtt{1},\mathtt{0},\mathtt{1}^{\hat{r}-1},\mathtt{1},\mathtt{1}), \quad
    \boldsymbol{f}_{\mathtt{1},7} := (\mathtt{0},\mathtt{0},\mathtt{1}^{\hat{r}-1},\mathtt{1},\mathtt{1}), \\
    \boldsymbol{f}_{\mathtt{1},8} &:= (\mathtt{1},\mathtt{1},\mathtt{1}^{\hat{r}-1},\mathtt{0},\mathtt{1}), \\
    \boldsymbol{f}_{\mathtt{0},6} &:= (\mathtt{0},\mathtt{0},\mathtt{0}^{\hat{r}-1},\mathtt{0},\mathtt{1}), \quad
    \boldsymbol{f}_{\mathtt{0},7} := (\mathtt{1},\mathtt{0},\mathtt{0}^{\hat{r}-1},\mathtt{0},\mathtt{1}), \\
    \boldsymbol{f}_{\mathtt{0},8} &:= (\mathtt{0},\mathtt{0},\mathtt{0}^{\hat{r}-1},\mathtt{1},\mathtt{1}).
  \end{align*}

  Table \ref{tab:1_seq} gives all the forbidden words and their mapping output $\rho(\boldsymbol{f}_{ij})$.
  In a similar way to the case of $z_m^{(\mathtt{0})} = z_m^{(\mathtt{1})} = \mathtt{0}$,
  we can obtain that the output $\vec{Z}$ of Algorithm \ref{alg:enc} is in $S_{n,r}$ if Eq.\ \eqref{eq:krd} holds.

  \begin{table}[t]
    \centering
    \caption{Forbidden words and mapping $\rho$}
    \label{tab:1_seq}
    \scalebox{.83}[.83]{
    \begin{tabular}{|c||ccccccc|c|}
      \hline
       & $p_1$ & $p_2$  &  $\cdots$  &  $p_{\hat{r}}$  & $p_{\hat{r}+1}$  &  $p_{\hat{r}+2}$ &  $p_m$ & $\rho(\boldsymbol{f}_{ij})$ \\
      $i$ & 1 &  2 & $\cdots$ & $\hat{r}$ & $\hat{r}+1$ & $\hat{r}+2$ & $m$ & \\ \hline
      $a_i$ & $2^0$ &  $2^1$  & $\cdots$ & $d_{\hat{r}}$ & $2^{\hat{r}-1}$ & $2^{\hat{r}}$ & $2^{\hat{r}}+1$  & \\ \hline
      $\boldsymbol{f}_{\mathtt{1},4}$ &  $\mathtt{1}$  &  $\mathtt{1}$  &  $\mathtt{1}^{\hat{r}-3}$  &  $\mathtt{1}$  &  $\mathtt{1}$  &  $\mathtt{1}$  &  $\mathtt{1}$  &  0\\
      $\boldsymbol{f}_{\mathtt{1},5}$ &  $\mathtt{0}$  &  $\mathtt{1}$  &  $\mathtt{1}^{\hat{r}-3}$  &  $\mathtt{1}$  &  $\mathtt{1}$  &  $\mathtt{1}$  &  $\mathtt{1}$  &  1\\
      $\boldsymbol{f}_{\mathtt{1},6}$ &  $\mathtt{1}$  &  $\mathtt{0}$  &  $\mathtt{1}^{\hat{r}-3}$  &  $\mathtt{1}$  &  $\mathtt{1}$  &  $\mathtt{1}$  &  $\mathtt{1}$  &  $2^{r}$ \\ \hline
      $\boldsymbol{f}_{\mathtt{1},7}$ &  $\mathtt{0}$  &  $\mathtt{0}$  &  $\mathtt{1}^{\hat{r}-3}$  &  $\mathtt{1}$  &  $\mathtt{1}$  &  $\mathtt{1}$  &  $\mathtt{1}$  &  $2^{\hat{r}+1}-1$ \\
      $\boldsymbol{f}_{\mathtt{1},8}$ &  $\mathtt{1}$  &  $\mathtt{1}$  &  $\mathtt{1}^{\hat{r}-3}$  &  $\mathtt{1}$  &  $\mathtt{1}$  &  $\mathtt{0}$  &  $\mathtt{1}$  &  $2^{\hat{r}+1}-2$ \\
      $\boldsymbol{f}_{\mathtt{0},6}$ &  $\mathtt{0}$  &  $\mathtt{0}$  &  $\mathtt{0}^{\hat{r}-3}$  &  $\mathtt{0}$  &  $\mathtt{0}$  &  $\mathtt{0}$  &  $\mathtt{1}$  &  $2^{\hat{r}+1}-3$ \\
      $\boldsymbol{f}_{\mathtt{0},7}$ &  $\mathtt{1}$  &  $\mathtt{0}$  &  $\mathtt{0}^{\hat{r}-3}$  &  $\mathtt{0}$  &  $\mathtt{0}$  &  $\mathtt{0}$  &  $\mathtt{1}$  &  $2^{\hat{r}+1}-4$ \\
      $\boldsymbol{f}_{\mathtt{0},8}$ &  $\mathtt{0}$  &  $\mathtt{0}$  &  $\mathtt{0}^{\hat{r}-3}$  &  $\mathtt{0}$  &  $\mathtt{0}$  &  $\mathtt{1}$  &  $\mathtt{1}$  &  $2^{\hat{r}}-1$ \\ \hline
    \end{tabular}
    }
  \end{table}

\end{IEEEproof}

\begin{re.}\upshape
  When $\hat{r} = 4$,
  we should set $d_{\hat{r}} = 6, 7$ from Theorem \ref{th:RLL_under}.
\end{re.}

The parameter $\hat{r}$ determines the sequence $\vec{a}$,
which are the coefficients of code constraint.
On the other hand, the parameter $r$ gives the maximum run-length.
Ordinary we set $r = \hat{r}$.
However, we can also set different values satisfying $r \geq \hat{r}$.
Theorem \ref{th:RLL} shows that Algorithm \ref{alg:enc} limits the maximum run-length when $r \geq \hat{r}$.

\begin{th.}\upshape
  \label{th:RLL}
  Suppose input $\boldsymbol{y}$ is in $S_{k,r}$.
  If $\hat{r} \geq 4$, $r \geq \hat{r}$, and $(k, r, d_{\hat{r}}) \not = (14,4,5)$,
  for all $b \in [0,a_{n+1}-1]$,
  the output $\boldsymbol{z}$ of Algorithm \ref{alg:enc} also satisfies $\boldsymbol{z} \in S_{n,r}$.
\end{th.}
\begin{IEEEproof}
  We have proven the statement in the case of $r = \hat{r}$.
  Hence,
  we should prove the statement for $r > \hat{r}$.
  The proof for $r = \hat{r}+1$ can be done in a similar way to $r = \hat{r}$.
  The proof for $r = \hat{r}+2$ is trivial
  since there does not exist any zero and one forbidden words.
\end{IEEEproof}

\subsection{Redundancy}
In this section, we compare the redundancy of the proposed code
and a lower bound of the redundancy of the optimal RLL-SIDC code.
For a code $\vec{C}$ of length $n$, 
we define the redundancy $\mathcal{R}(\vec{C})$ as follows:
\begin{align*}
  \mathcal{R}(\vec{C}) := n - \log_2 |\vec{C}|,
\end{align*}
where $|\vec{C}|$ represents the cardinality of a code $\vec{C}$.
Roughly speaking, the redundancy is the number of additional symbols to encode a message.

Kulkarni and Kiyavash \cite{kulkarni2013nonasymptotic} presented
an upper bound of the cardinality of the optimal SIDC code $\vec{C}_{\mathbf{opt}} (n)$ as $|\vec{C}_{\mathbf{opt}} (n)| \leq \frac{2^n -2}{n-1}$.
Here, the optimal code means the code with the largest cardinality.
Moreover, the cardinality of the optimal RLL-SIDC code $\vec{C}_{\mathbf{opt}}^{\mathbf{RLL}} (n)$ is
less than or equal to one of the optimal SIDC code,
i.e., $|\vec{C}_{\mathbf{opt}}^{\mathbf{RLL}} (n)| \leq |\vec{C}_{\mathbf{opt}} (n)|$.
Hence, we get
\begin{align*}
  |\vec{C}_{\mathbf{opt}}^{\mathbf{RLL}} (n)| 
  \leq 
  \frac{2^n -2}{n-1}.
\end{align*}
This leads a lower bound of the redundancy of the optimal RLL-SIDC code as follows:
\begin{align}
  \label{eq:redundancy_opt}
  \mathcal{R}(\vec{C}_{\mathbf{opt}}^{\mathbf{RLL}} (n))
  &\geq
  n - \log_2 (2^n - 2) + \log_2 (n - 1) \nonumber \\ 
  &=: \phi (n).
\end{align}

Let us evaluate the redundancy of the code derived from the proposed encoding algorithm in Sect.\ \ref{sec:encoding}.
Recall that once the code length $n$ is fixed,
the parameter $\hat{r}$ is decided by Algorithm \ref{alg:enc}.
More precisely, $\hat{r}$ becomes 
the smallest positive integer satisfying $n \leq 2^{\hat{r}} + \hat{r} + 1$.
Hence, hereafter, we denote the proposed code, by $\vec{C}_b(n)$, to simplify the notation.

Firstly, we will evaluate the redundancy by using parameter $\hat{r}$.
As shown in Fig.\ \ref{fig:ov}, the length of message (resp.\ codeword) is $k-1$ (resp.\ $n$).
Hence, the redundancy is $n-k+1$.
Combining this and Eqs.\ \eqref{eq:m}, and \eqref{eq:n}, we get
\begin{align}
  \label{eq:redundancy}
  \mathcal{R}(\vec{C}_b(n))
  = 
  \hat{r} + 4.
\end{align}

Secondly, we will evaluate the code length $n$ by using parameter $\hat{r}$.
Equation \eqref{eq:hat_r} leads
\begin{align*}
  k \in [2^{\hat{r}-1} - 1, 2^{\hat{r}} - 2].
\end{align*}
Combining this condition and Eqs.\ \eqref{eq:m} and \eqref{eq:n}, we get
\begin{align}
  \label{eq:n_range}
  n \in [2^{\hat{r}-1} + \hat{r} + 2, 2^{\hat{r}} + \hat{r} + 1] =: L_{\hat{r}}.
\end{align}
From Eqs.\ \eqref{eq:redundancy} and \eqref{eq:n_range}, we obtain the relationship between redundancy and code length.

Theorem \ref{th:redundancy} shows the difference 
between the redundancy of the proposed code 
and the lower bound of the redundancy of the optimal RLL-SIDC code.

\begin{th.}\upshape
  \label{th:redundancy}
  For $n \geq \underline{L_{4}} = 14$, 
  \begin{align*}
    \mathcal{R}(\vec{C}_b (n)) - \phi (n) < 5.
  \end{align*}
  In words,
  the difference between the redundancy of the proposed code 
  and a lower bound of the redundancy of the optimal RLL-SIDC code
  is less than $5$.
\end{th.}

We show a lemma required for proving Theorem \ref{th:redundancy}.

\begin{le.}\upshape
  \label{le:phi}
  Define $\phi (n)$ as in Eq.\ \eqref{eq:redundancy_opt}.
  For $n \geq 14$,
  $\phi (n)$ is the monotonically increasing.
\end{le.}

\begin{IEEEproof}[Proof of Lemma \ref{le:phi}]
  To prove Lemma \ref{le:phi}, 
  we show that the derived function $\frac{\phi (n)}{dn}$ is always positive.
  The derived function $\frac{\phi (n)}{dn}$ is
  \begin{align}
    \label{eq:d_phi}
    \frac{d \phi(n)}{d n}
    &=
    \frac{2^n - 2 - 2(n-1) \log_{e}2}{(2^n-2)(n-1)\log_{e}2} \nonumber \\
    &=:
    \frac{\psi (n)}{(2^n-2)(n-1)\log_{e}2}.
  \end{align}
  Note that $\psi (14) > 0$.
  For $n \geq 14$,
  \begin{align*}
    \psi (n+1) - \psi (n)
    =
    2^n - 2 \log_{e}2
    >
    0,
  \end{align*}
  holds.
  Hence, function $\psi (n)$ is always positive.
  Therefore, for $n \geq 14$, Eq.\ \eqref{eq:d_phi} is always positive.
\end{IEEEproof}

\begin{figure}[t]
  \begin{center}
    \includegraphics[width=.7\linewidth]{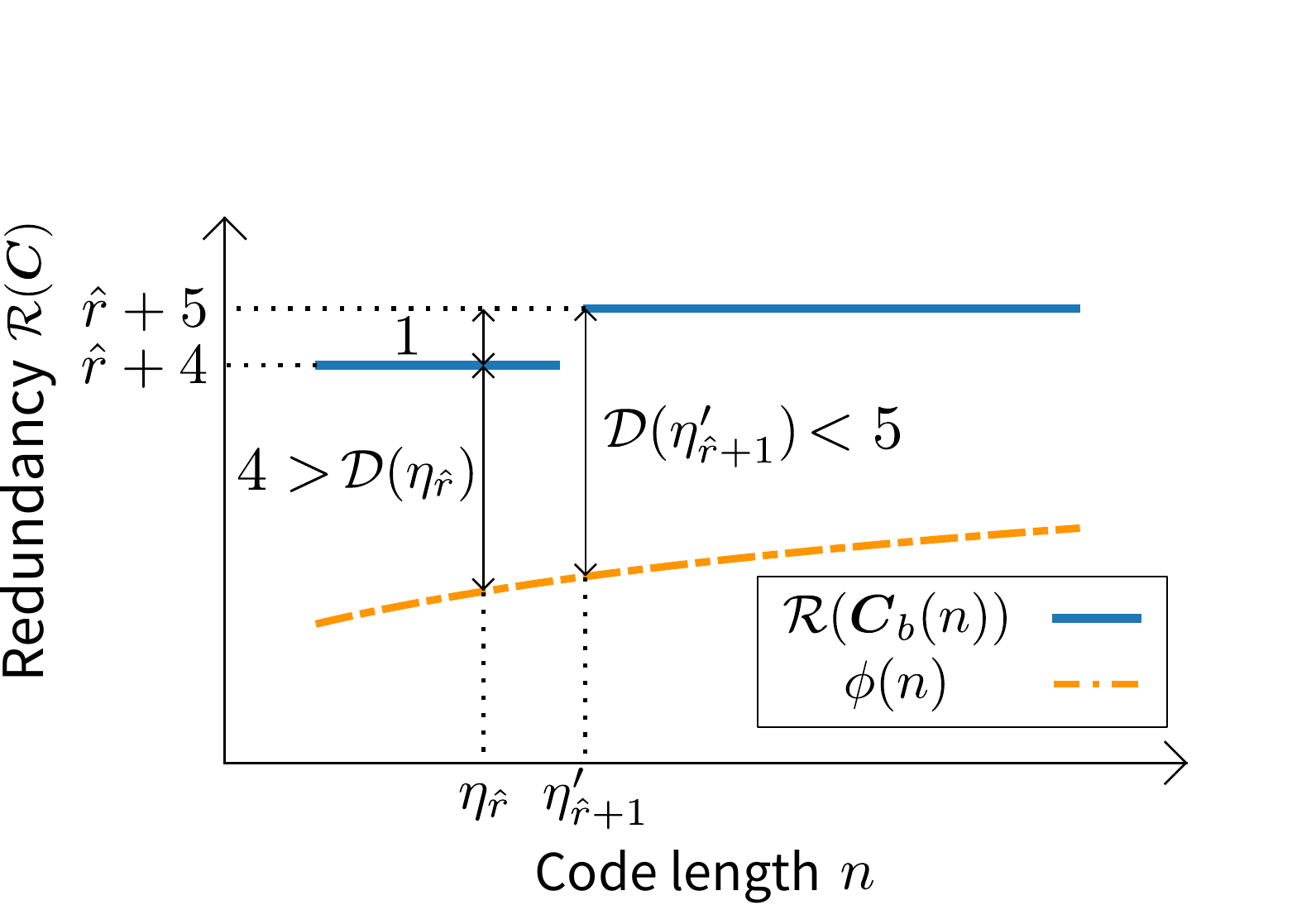}
    \caption{Outline of the difference of the redundancy}
    \label{fig:redundancy}
  \end{center}
\end{figure}

Figure \ref{fig:redundancy} depicts the outline of proof of Theorem \ref{th:redundancy}.
Firstly, for a fixed $\hat{r}$, we will show that 
an upper bound of the difference at $n = 2^{\hat{r}} + 1$ is less than $4$.
Secondly, we will show  
$\mathcal{R}(\vec{C}_b) - \phi (n) < 5$.

\begin{IEEEproof}[Proof of Theorem \ref{th:redundancy}]
  Define
  \begin{align}
    \label{eq:Dn}
    \mathcal{D} (n) := \mathcal{R}(\vec{C}_b(n)) - \phi (n).
  \end{align}
  Denote
  \begin{align}
    \label{eq:eta}
    \eta_{\hat{r}} = 2^{\hat{r}} + 1,
  \end{align}
  for a positive integer $\hat{r}$.
  Note that $\eta_{\hat{r}} \in L_{\hat{r}}$.
  Equation \eqref{eq:eta} leads $\hat{r} = \log_2 (\eta_{\hat{r}}-1)$.
  Firstly, we evaluate $\mathcal{D} (\eta_{\hat{r}})$ for $\hat{r} \geq 4$.
  From Eq.\ \eqref{eq:redundancy}, 
  the redundancy of the proposed encoding algorithm is as follows:
  \begin{align}
    \label{eq:rc_eta}
    \mathcal{R}(\vec{C}_b(\eta_{\hat{r}}))
    = 
    \log_2(\eta_{\hat{r}}+1) + 4.
  \end{align}
  From Eqs.\ \eqref{eq:redundancy_opt}, \eqref{eq:Dn}, and \eqref{eq:rc_eta},
  we get 
  \begin{align*}
    \mathcal{D} (\eta_{\hat{r}})
    =
    -\eta_{\hat{r}} + \log_2(2^{\eta_{\hat{r}}} - 2) + 4
    <
    4.
  \end{align*}

  Secondly, for a fixed $\hat{r}$, we discuss the maximum value of $\mathcal{D} (n)$.
  From Lemma \ref{le:phi} and Eq.\ \eqref{eq:redundancy}, 
  for $n \in L_{\hat{r}}$, $\mathcal{D} (n)$ attains its maximum value at
  \begin{align*}
    n
    =
    \underline{L_{\hat{r}}}
    =:
    \eta_{\hat{r}}'.
  \end{align*}
  So, we will calculate $\mathcal{D} (\eta_{\hat{r}}')$.
  From Eq.\ \eqref{eq:redundancy}, 
  $\mathcal{R}(\vec{C}_b(\eta_{\hat{r}+1}')) = \mathcal{R}(\vec{C}_b(\eta_{\hat{r}})) + 1$ holds.
  Moreover, for $\hat{r} \geq 4$, $\eta_{\hat{r}} < \eta_{\hat{r}+1}'$ holds.
  Hence, we obtain $\phi (\eta_{\hat{r}}) < \phi (\eta_{\hat{r}+1}')$.
  Combining these, for $\hat{r} + 1 \geq 5$, we get
  \begin{align*}
    \mathcal{D} (\eta_{\hat{r}+1}')
    &=
    \mathcal{R}(\vec{C}_b(\eta_{\hat{r}+1}')) - \phi (\eta_{\hat{r}+1}')  \\
    &=
    \mathcal{R}(\vec{C}_b(\eta_{\hat{r}})) - \phi (\eta_{\hat{r}+1}') + 1  \\
    &<
    \mathcal{R}(\vec{C}_b(\eta_{\hat{r}})) - \phi (\eta_{\hat{r}}) + 1  \\
    &=
    \mathcal{D} (\eta_{\hat{r}}) + 1  \\
    &<
    5.
  \end{align*}
  For $\hat{r} = 4$, we get $\mathcal{D} (\eta_{4}') \approx 4.299 < 5$.
  Thus, for $n \geq 14$,
  $\mathcal{R}(\vec{C}_b (n)) - \phi (n) < 5$ holds.
\end{IEEEproof}

\section{Conclusion}
In this paper,
we compare the RLL sequence encoder by the WI algorithm and the NRZI with the one by Schoeny et al.\ \cite[Appendix B]{schoeny2017codes}.
we proposed an SIDC code which is easily limited the maximum run-length
and an encoding algorithm for it.
Moreover,
we proved that the maximum run-length of the output of the algorithm is limited.
Furthermore, we compare the redundancy of the proposed encoding algorithm and the lower bound of the redundancy of the optimal RLL-SIDC code.

\section*{Acknowledgment}
We would like to express my gratitude to Dr.\ Hagiwara at Chiba University
to introduce monotonically increasing codes.
This research was supported by
Inamori Research Grants
and
Yamaguchi University Fund.

\end{document}